# The Importance of Short- and Long-Range Exchange on Various Excited State Properties of DNA Monomers, Stacked Complexes, and Watson-Crick Pairs

Alexandra E. Raeber and Bryan M. Wong*

Department of Chemical & Environmental Engineering and Materials Science & Engineering Program, University of California, Riverside, Riverside, CA 92521, USA

*Corresponding author. E-mail: bryan.wong@ucr.edu. Homepage: http://www.bmwong-group.com

**Abstract**

We present a detailed analysis of several time-dependent DFT (TD-DFT) methods, including conventional hybrid functionals and two types of non-empirically tuned range-separated functionals, for predicting a diverse set of electronic excitations in DNA nucleobase monomers and dimers. This large and extensive set of excitations comprises a total of 50 different transitions (for *each* tested DFT functional) that includes several n $\to \pi$ and $\pi \to \pi^*$ valence excitations, long-range charge-transfer excitations, and extended Rydberg transitions (complete with benchmark calculations from high-level EOM-CCSD(T) methods). The presence of localized valence excitations as well as extreme long-range charge-transfer excitations in these systems poses a serious challenge for TD-DFT methods that allows us to assess the importance of both short- and long-range exchange contributions for simultaneously predicting all of these various transitions. In particular, we find that functionals that do not have *both* short- and full long-range exchange components are unable to predict the different types of nucleobase excitations with the same accuracy. Most importantly, the current study highlights the importance of both short-range exchange and a non-empirically tuned contribution of long-range exchange for accurately predicting the diverse excitations in these challenging nucleobase systems.

## 1. Introduction

The electronic properties of DNA nucleobase complexes continue to be an active area of research due to their importance in condensed-phase chemistry,[1-3] nanotechnology,[4, 5] and new bio-detection technologies.[6] In particular, a deep understanding of nucleobase complexes using first-principles methods is vital for these new technologies since electronic effects directly impact the stability[7] and optical properties[8] of nanostructures that are assembled from these molecules. As researchers continue to use DNA complexes to create three-dimensional nanostructures for circuits and plasmonic devices,[9-12] there is a crucial need for efficient *and* accurate theoretical methods for predicting the electronic properties of these large systems.

Despite the growing amount of experimental and theoretical work in this field, the electronic properties of DNA double helices are still not well understood and remain controversial.[13-16] Even at nucleobase monomer and dimer sizes, discrepancies between experiment and the different theoretical models remain, and the path to improving the theoretical predictions for these systems is not obvious. As a result, a detailed study of computationally-efficient density functional theory (DFT) methods against high-quality wavefunction-based models is necessary for obtaining efficient and accurate predictive methods for these complex systems. Very recently, Szalay and co-workers[17] presented a thorough study of excitation energies of all the DNA nucleobases (adenine, cytosine, guanine, and thymine) as well as the stacked adenine-thymine pair, stacked guanine-cytosine, and the Watson-Crick (WC) pair of guanine-thymine, as shown in Fig. 1. In this previous study, Szalay and co-workers calculated several excited states including valence excitations, charge-transfer (CT) transitions, and Rydberg excitations using the equation of motion coupled-cluster method with single and double excitations (EOM-CCSD). In particular, these researchers found that the inclusion of perturbative triple excitations at the EOM-CCSD(T) level of theory was essential since the errors of lower-level EOM-CCSD methods could be as large as 0.3 eV. At the TD-DFT level of theory, we are fully aware of previous studies on WC pairs,[18, 19] hydrogen-bonded and stacked complexes,[20, 21] adenine dimers,[22, 23] cytosine dimers,[24] thymine dimers,[25] and AT and GC pairs.[26, 27] However, as pointed out by Szalay et al., most of these previous studies only investigated a few low-lying excitation states, and comparisons in these studies were typically made to lower-level wavefunction-based methods such as CIS, CC2, and EOM-CCSD (without the important perturbative triple excitations). Considering the high computational cost of the EOM-CCSD(T) calculations, Szalay et al. specifically called for "new approximate methods that can treat different types of excitations with the same accuracy" in these systems.[17]

In this work, we present a detailed analysis of several modern time-dependent DFT (TD-DFT) methods benchmarked against the high-level EOM-CCSD(T) calculations by Szalay and co-workers. The DFT methods studied here include several conventional hybrid functionals as well as two types of range-separated functionals that have been non-empirically tuned to satisfy Janak's theorem.[28] As explained extensively in Ref. 29, Koopman's theorem allows us to equate the fundamental gap to the band energy difference in Hartree-Fock theory; however, in Kohn-Sham DFT, one must instead use Janak's theorem as it accounts for the discontinuity in the exchange-correlation potential in an $N$-electron system. Our group and other researchers have previously used these range-separated functionals to predict long-range charge transfer effects in solar cell dyes,[30, 31] quasiparticle gaps in molecules,[29, 32] and excitation states in light-harvesting organic photovoltaics.[33-37] In this study on DNA nucleobase complexes, we examine a large and diverse set of excitations that comprises a total of 50 different transitions, including several n → π and π → π* valence excitations, long-range charge-transfer excitations, and extended Rydberg transitions. Most importantly, in order to address the need for treating different types of excitations with the same accuracy, we specifically highlight the importance of using *both* short- and full long-range exchange for accurately predicting the various excitations in these complex nucleobases. Specifically, we find that functionals that *do not* have both short- and full long-range exchange components are unable to predict the different types of excitations in these systems with the same accuracy. Finally, we give a detailed analysis for each of the nucleobases and discuss the implications for simultaneously predicting the diverse excitations in these challenging nucleobase systems.

## 2. Theory and Methodology

Since the purpose of this work is to assess the accuracy of both conventional and range-separated functionals (with particular emphasis on the short- and long-range exchange contributions in the latter) in various nucleobase excited states, we briefly outline the underlying theories for each. One of the most widely used linear-response TD-DFT schemes for calculating excitation energies involves the use of global hybrid functionals, where a certain fraction of Hartree-Fock (HF) exchange is admixed with a DFT approximation for the exchange-correlation energy. Within a simple one-parameter mixing scheme, the exchange-correlation energy for a conventional global hybrid functional is given by

$$E_{xc}^{\text{global}} = \alpha E_{x,\text{HF}} + (1-\alpha) E_{x,\text{DFT}} + E_{c,\text{DFT}}, \qquad (1)$$

where $E_{x,\text{HF}}$ is the HF exchange energy based on Kohn-Sham orbitals, $E_{x,\text{DFT}}$ is a DFT contribution to the exchange energy, and $E_{c,\text{DFT}}$ is the correlation functional due to DFT. In the literature, there are numerous hybrid functionals that combine different DFT treatments of exchange and correlation with varying coefficients. In addition, there are several other hybrid functionals that utilize more than one parameter for mixing DFT and HF exchange contributions (i.e., Becke's popular B3LYP method[38] utilizes a three-parameter mixing scheme). Despite their different parameterizations, both the B3LYP and M06-HF[39, 40] functionals used in this work are categorized as "global hybrid" functionals since the fraction of nonlocal HF exchange, $\alpha$, is held constant in space and fixed to a specific value. The B3LYP functional, for example, is parameterized with $\alpha = 0.20$, and the M06-HF functional is parameterized with $\alpha = 1.0$. We have chosen these specific functionals for comparison due to their wide-spread use and because they represent two different extremes of global hybrids where the HF exchange contribution ranges from 0.20 to 1.00.

In contrast to conventional hybrid functionals, the range-separated formalism[41, 42] mixes short range density functional exchange with long range Hartree-Fock exchange by separating the electron repulsion operator into short and long range terms (i.e., the mixing parameter is a function of electron coordinates). In its most general form, the partitioning is given by

$$\frac{1}{r_{12}} = \frac{1-[\alpha + \beta \cdot \text{erf}(\mu \cdot r_{12})]}{r_{12}} + \frac{\alpha + \beta \cdot \text{erf}(\mu \cdot r_{12})}{r_{12}}. \quad (2)$$

The "erf" term denotes the standard error function, $r_{12}$ is the interelectronic distance between electrons 1 and 2, and $\mu$ is the range-separation parameter in units of Bohr$^{-1}$. The other extra parameters, $\alpha$ and $\beta$, satisfy the following inequalities: $0 \leq \alpha + \beta \leq 1$, $0 \leq \alpha \leq 1$, and $0 \leq \beta \leq 1$. The parameter $\alpha$ in the partitioning allows a contribution of HF exchange over the entire range by a factor of $\alpha$, and the parameter $\beta$ allows us to incorporate long-range asymptotic HF exchange by a factor of $(\alpha + \beta)$. When $\alpha = 0.2$ and $\beta = 0.0$, the exchange-correlation energy reduces to a B3LYP-like functional (as shown in Ref. [43], it is not exactly B3LYP due to an extra exchange term, but the two expressions are closely related). The CAM-B3LYP functional of Yanai and co-workers[44] uses $\alpha = 0.19$, $\alpha + \beta = 0.65$, and $\mu = 0.33$; however, the CAM-B3LYP functional *does not* incorporate a "full" range-separation as it only has 65% HF exchange at long-range (instead of the correct 100% asymptotic HF exchange). In our previous work on range-separated functionals,[30, 32, 33, 36, 45] we have used and parameterized "full" range-separation schemes that correspond to setting $\alpha = 0.0$ and $\beta = 1.0$. In particular, we[33] and others [34, 35] have previously shown that maintaining a full 100% contribution of asymptotic HF exchange is essential for accurately describing valence excitations in even relatively simple molecular systems. However, there has been recent work[46-49] suggesting that some amount of short-range HF exchange (i.e. setting $\alpha$ to a nonzero value) can lead to improved electronic properties and excitation energies. For the two range-separated methods used in this work, we fix $\alpha + \beta = 1.0$ (with different values of $\alpha$) in conjunction with tuning the range-separation parameter $\mu$ via the non-empirical procedure by Baer and Kronik[29, 31, 50] discussed below. In short, the presence of the two extra parameters, $\alpha$ and $\beta$, in Equation (2) gives us extra flexibility in assessing the importance of both short- and long-range exchange for *simultaneously* predicting all the diverse transition energies and properties in various nucleobase complexes.

For a full long-range corrected functional with given values of $\alpha$ and $\beta$ (such that $\alpha + \beta = 1.0$), Baer and Kronik[29, 31, 50] have demonstrated that the range-separation parameter can be tuned non-empirically by (approximately) satisfying Janak's theorem. In a molecular system, this is done by ensuring that the ionization potential (IP) and the negative of the HOMO energy for the *N* electron system are equal. The exact exchange-correlation functional would automatically satisfy this condition, providing theoretical justification for self-consistently tuning $\mu$ within this procedure. The difference between the ground state energy of the *N* electron and the *N* – 1 electron system gives its IP, which according to Janak's theorem, is equal to the negative of the HOMO energy, $\varepsilon_{HOMO}(N)$. A range-separation parameter that approximately satisfies this condition can be obtained by minimization of the objective function

$$J^2(\mu) = \left[\varepsilon^{\mu}_{HOMO}(N) + IP^{\mu}(N)\right]^2 + \left[\varepsilon^{\mu}_{HOMO}(N+1) + IP^{\mu}(N+1)\right]^2. \quad (3)$$

where $\varepsilon^{\mu}_{HOMO}(N)$ is the HOMO of the *N*-electron system, and $IP^{\mu}(N)$ is the energy difference between the ground state energies of the *N* and *N* – 1 electron systems *with the same value of μ.* As mentioned previously and described in Ref. 29, the derivative of the total DFT energy with respect to electron number is discontinuous at the *N*-electron point, and a theorem that formally relates the LUMO energy to the electron affinity does not exist. This problem is circumvented by including the second term in Eq. (3) by considering the HOMO of the *N* + 1 electron system. Although this nonempirical tuning procedure directly modifies the HOMO and LUMO energies of the system, we[32] and others[29] have previously shown that this method also significantly improves the description of excited state properties, which we explore further in this study.

In order to accurately compare our TD-DFT calculations to high-level EOM-CCSD(T) benchmarks, identical molecular geometries obtained from Szalay et al.[17] were used in this work. The Cartesian coordinates for all the systems studied here are listed in the Supporting Information for completeness. In addition, and most crucial to our study, difference densities for all of our TD-DFT calculations (including all 50 transitions for each DFT functional) were generated and carefully compared to the original transition densities from Szalay et al.[17] Optimal *μ* values were determined for adenine, cytosine, guanine, thymine, an adenine-thymine stacked pair, a guanine-cytosine stacked pair, and a guanine-cytosine Watson-Crick pair. For each of these systems, we computed $J^2$ in Equation (3) with the TZVP basis set using two different parameterizations: a long-range corrected BLYP (LC-BLYP) functional without any short-range exchange (i.e., *α* = 0.0, *β* = 1.0) as well as an LC-BLYP functional containing 20% exchange over the entire range (i.e., $\alpha = 0.2$, $\beta = 0.8$). The choice of $\alpha = 0.2$ in the latter is motivated by a very recent study by Kronik et al.[48] which found that (non-empirically tuned) values of $\alpha \sim 0.2$ in conjunction with long-range exchange were able to accurately predict outer-valence electron spectra of various heterocyclic systems. The two-dimensional tuning procedure by Autschbach[51] provides a non-empirical prescription for determining $\alpha$; however, the effort required to assign all 50 nucleobase excitation energies (for *each* DFT method) is already quite onerous, so we reserve these two-dimensional tuning approaches for a subsequent study. Nevertheless, it is important to note that both of the different LC-BLYP parameterizations used in this work still recover the full 100% exchange at asymptotic distance ($\alpha + \beta = 1.0$) even though each parameterization has a different exchange contribution at short range. In order to determine the optimal range-separation value for each system, we carried out several single-point energy calculations by varying $\mu$ from 0.05 to 0.7 (in increments of 0.05) for each of the $N$, $N + 1$, and $N - 1$ electron states. Figure 2 shows the smooth curves resulting from computing $J^2$ as a function of $\mu$ for each of the nucleobase geometries. Spline interpolation was used to refine the minimum for each individual system, and Table 1 contains a summary of the optimal $\mu$ values. It is worth noting that the short-range DFT exchange in Eq. (2) decays rapidly on a length scale of $\sim 1/\mu$ and, therefore, smaller values of $\mu$ are more appropriate for larger molecules (i.e., a smaller value of $\mu$ enables the short-range Coulomb operator to fully decay to zero on the length scale of the molecule). Indeed, the optimal $\mu$ values in Table 1 reflect these trends with the larger-sized dimers having slightly smaller values of $\mu$ than the monomers. Once they were determined, the $\mu$ parameters given in the table were used for all subsequent LC-BLYP TD-DFT calculations. All calculations were carried out with the Gaussian 09 package[52] using default SCF convergence criteria (density matrix converged to at least $10^{-8}$) and the default DFT integration grid (75 radial and 302 angular quadrature points). For each system, the 20 lowest excitation states were determined and assigned by examining both the oscillator strength and the charge density difference between the ground and excited states and comparing them to the assignments given in Szalay et al.[17] Rydberg orbitals are denoted by R, pi orbitals by $\pi$, and lone pair orbitals by n. Virtual orbitals are designated by an asterisk (*) with a preceding number referring to its orbital number. Visualizations of the charge density differences other than those given in the figures of the main text are provided in the Supporting Information.

While charge transfer excitations are indicated by the "CT" transition assignments in the tables (as originally assigned by Szalay et al.), we also used the lambda diagnostic by Peach et al.[53] to give a more quantitative measure of charge transfer for all of the calculated excitation energies. The spatial overlap between the occupied and virtual orbitals involved in an excitation, $\Lambda$, is quantified in the form

$$\Lambda = \frac{\sum_{i,a}(X_{ia}+Y_{ia})^2 O_{ia}}{\sum_{i,a}(X_{ia}+Y_{ia})^2}, \qquad (4)$$

where $X_{ia}$ and $Y_{ia}$ are the virtual-occupied and occupied-virtual transition amplitudes, respectively, and $O_{ia}$ is the spatial overlap of the moduli of the two orbitals. By construction, $\Lambda$ is bounded between 0 and 1, with small values signifying a long-range charge-transfer excitation and large values signifying a localized, short-range transition. Extensive benchmarks given by Peach and co-workers indicated that excitations with $\Lambda < 0.3$ imply little orbital overlap and significant long-range charge transfer excitations that produce inaccuracies in hybrid functionals.[53] The $\Lambda$ diagnostic was carried out for all systems and their various excitation energies at the B3LYP/TZVP level of theory.

**3. Results and Discussion**

A concise summary and analysis of all 50 nucleobase excited-states obtained by TD-DFT (M06-HF, B3LYP, CAM-B3LYP, LC-BLYP$_{\alpha=0.0,\beta=1.0}$ and LC-BLYP$_{\alpha=0.2,\beta=0.8}$) in comparison to EOM-CCSD(T) benchmarks is given in Table 2. Detailed excited state energies, oscillator strengths, and transition assignments are given in Table 3 for the monomers, and similar data are collected in Table 4 for the stacked pairs and Watson-Crick pair. Overall, the range-separated functionals give significantly better predictions for the excited state properties than the conventional hybrid functionals. Based on the mean absolute error, the predictions made by CAM-B3LYP and the two LC-BLYP functionals are roughly equivalent for the monomers; however, both versions of LC-BLYP are considerably more accurate than CAM-B3LYP for the GC Watson-Crick pair, particularly for CT excitations in this system. We discuss and analyze the results in detail for each of the various systems in the following sections.

**Monomers**

Upon examination of Table 3, we find that all five TD-DFT methods give the correct ordering of transitions for adenine, cytosine, guanine, and thymine. In this work, the excitation energies for adenine are most accurately calculated by LC-BLYP$_{\alpha=0.2,\beta=0.8}$ with a MAE of 0.09 and least accurately predicted by B3LYP. In the case of thymine, B3LYP has the highest MAE and RMS errors (0.46 and 0.48, respectively) and CAM-B3LYP has the lowest (0.08 and 0.10), though the other range-separated functionals also have comparatively low MAE and RMS values. The results for guanine are similar with B3LYP and M06-HF both having MAE values of approximately 0.4 and the best of the range-separated functionals (LC-BLYP$_{\alpha=0.2,\beta=0.8}$) decreasing the error tenfold. It is interesting to note that all the TD-DFT excited state energies for cytosine are slightly less accurate than the other nucleobases, with the highest MAE and RMS values (0.56 and 0.72, respectively) for M06-HF and the lowest (0.08 and 0.095) from LC-BLYP$_{\alpha=0.0,\beta=1.0}$.

For both guanine and thymine, Rydberg excitations are also present and were included in the EOM-CCSD(T)/TZVP benchmark calculations.[17] We are acutely aware that diffuse basis functions are required to accurately describe Rydberg states and, indeed, Szalay et al. have reported excitation energies of nucleobases that properly treat Rydberg states with augmented basis sets in a series of previous papers.[54, 55] In order to make a fair comparison to the most recent EOM-CCSD(T)/TZVP benchmarks of Szalay et al.,[17] we have chosen to use the TZVP basis throughout this work. A more thorough analysis of Rydberg excitations would require a detailed comparison to Szalay's previous benchmark calculations,[54, 55] which we reserve for a future study. For these Rydberg transitions, the excitations are especially sensitive to the asymptotic part of the nonlocal exchange-correlation potential, and one should expect that long-range exchange corrections are important. Indeed, we find that the B3LYP hybrid functional dramatically underestimates these Rydberg excitations while all three range-separated functionals (and even M06-HF) compare extremely well with the EOM-CCSD(T) benchmarks.

In general, the MAE for the conventional hybrid functionals have similar values of approximately 0.4, while LC-BLYP$_{\alpha=0.2,\beta=0.8}$, which includes some short-range exchange, has an overall MAE of approximately 0.1 – a noticeable improvement in accuracy. In particular, the M06-HF functional is the least accurate method overall and significantly overestimates the excitations energies for all of the nucleobase monomers. This overestimation is not surprising since the majority of the monomer excitations in Table 3 are valence transitions and do not have long-range intramolecular charge-transfer character, as indicated by values of $\Lambda > 0.4$. Since the M06-HF functional is parameterized with a full 100% HF exchange over the entire range ($\alpha = 1.0$ over *all* space), the intricate balance between exchange and correlation errors is corrupted for these valence excitations, resulting in excitation energies that are severely overestimated. Although the valence excitations only involve localized short-range transitions, we still find that some amount of short-range HF exchange (present in both CAM-B3LYP and LC-BLYP$_{\alpha=0.2,\beta=0.8}$) slightly improves the accuracy of these excitations in comparison to the standard LC-BLYP$_{\alpha=0.0,\beta=1.0}$ results. We posit that non-negligible self-interaction errors (SIE) are still present in these short-range transitions and the inclusion of some short-range HF exchange partially reduces the SIE in these localized excitations. As a whole, we find that both the CAM-B3LYP and LC-BLYP$_{\alpha=0.2,\beta=0.8}$ functionals give an accurate and balanced prediction of the various valence and Rydberg excitations in the nucleobase monomers.

**Adenine-Thymine Stacked Pair**

We now turn to our first nucleobase dimer composed of an adenine and thymine $\pi$-stacked geometry with an intermolecular separation of 3.154 Å. The excitation energies and oscillator strengths for the AT stacked pair are given in Table 4, and a visualization of the charge density difference for various excitations is given in Figure 3. As seen in the table, all three range-separated methods more closely reproduce the EOM-CCSD(T) benchmark excited state energies and oscillator strengths, compared to M06-HF or the widely-used B3LYP functional. While all three perform equally well for the majority of the excitation energies, the standard LC-BLYP$_{\alpha=0.0,\beta=1.0}$ functional is the most accurate for the CT$_{A\rightarrow T}$ excitation originally assigned by Szalay et al. For this specific excitation, CAM-B3LYP has the highest error among the range-separated functionals, followed by LC-BLYP$_{\alpha=0.2,\beta=0.8}$. It is particularly interesting that we also find several excitations in the AT stacked pair that have $\Lambda < 0.4$, indicating charge-transfer, even though none of these are explicitly assigned as charge-transfer excitations in the Szalay paper (i.e., $A_{n-1} \rightarrow 2\pi^*$ and $T_{n-1} \rightarrow 2\pi^*$). For these particular excitations, the range-separated methods that include short-range exchange (CAM-B3LYP and LC-BLYP$_{\alpha=0.2,\beta=0.8}$) give the best predictions, with the standard LC-BLYP$_{\alpha=0.0,\beta=1.0}$ being the least accurate.

As a whole, the TD-DFT methods give reasonable predictions for the specific ordering of excited-state energies, although there are a few outliers. All of the techniques place the $1(\pi\pi^*)$ transition higher in energy than the $2(\pi\pi^*)$ transition, breaking the smooth increasing trend in the benchmark data; however, this effect is much worse for M06-HF than for the other four methods. In the EOM-CCSD(T) benchmark data, the $4(\pi\pi^*)$ transition is lower in energy than the $3(n\pi^*)$ transition, but the opposite is true for all of the TD-DFT methods except B3LYP. In addition, all methods switch the energetic ordering of $5(\pi\pi^*)$ and $5(n\pi^*)$ as compared to the EOM-CCSD(T) benchmark data. Turning to the Rydberg excitations in the AT stacked pair, we find that all three range-separated functionals (as well as M06-HF) describe these excitations accurately. The conventional B3LYP functional performs quite poorly by severely underestimating this excitation by 0.56 eV. Among all the range-separated functionals, CAM-B3LYP has the smallest error for the Rydberg excitations by a small margin. However, the LC-BLYP$_{\alpha=0.2,\beta=0.8}$ functional gives an overall more accurate and balanced prediction of the various valence and Rydberg excitations in the AT stacked pair.

**Guanine-Cytosine Stacked Pair**

Our next nucleobase dimer is another $\pi$-stacked system, but composed of a guanine and cytosine geometry separated by a distance of 3.104 Å. The excitation energies and oscillator strengths are given in Table 4, and visualizations of the charge density differences for selected excitations are given in Figure 4. Once again, all three range-separated functionals more closely reproduce the EOM-CCSD(T) benchmark excited state energies and oscillator strengths, compared to M06-HF or B3LYP. Although all three have similar overall levels of error, the energies of the two charge transfer excitations are most accurately predicted by the standard LC-BLYP$_{\alpha=0.0,\beta=1.0}$ functional, as in the AT stacked pair. For the GC stacked pair, $\Lambda$ values of around 0.3, however, do occur for the excitations that are assigned as charge transfers, giving further weight to the assignment. As these excitations also have $\pi \rightarrow \pi^*$ character as well as charge transfer, the $\Lambda$ values are not as low as they would be for a full charge transfer. Overall, the oscillator strengths are well predicted by the TD-DFT methods, though they are least accurate for B3LYP. The Rydberg excitations are well described by the range-separated functionals with CAM-B3LYP having the lowest error by a small margin. The M06-HF functional, which incorporates full 100% asymptotic HF exchange, described the Rydberg transitions less well, and the B3LYP functional describes these same excitations quite poorly.

As is the case for the AT stacked pair, the ordering of excited-state energies in the GC stack relative to the EOM-CCSD(T) benchmarks is relatively good with a few outliers, although M06-HF has one more error than the other four methods. Specifically, the 2(n$\pi^*$) transition is placed at a lower energy than the 1(n$\pi^*$) transition in M06-HF, while the opposite is true in the benchmark data. While the EOM-CCSD(T) data places the 3($\pi\pi^*$) charge-transfer transition at a lower energy than the 4($\pi\pi^*$) transition, the TD-DFT methods all show the opposite trend. The opposite problem occurs with the 3(n$\pi^*$) transition, which is placed at a lower energy than the 5($\pi\pi^*$) transition by the TD-DFT methods but at a higher energy by EOM-CCSD(T). The inconsistencies in transition order for this second pair, however, do not involve especially high absolute errors for the range-separated functionals. Overall, the LC-BLYP$_{α=0.0,β=1.0}$ functional (without short-range exchange) best describes the excitation energies of the guanine-cytosine stacked pair, with the other two range separated functionals performing nearly as well.

**Guanine Cytosine Watson-Crick Base Pair**

Finally, we examine a Watson-Crick (WC) nucleobase pair formed via 3 hydrogen bonds between a guanine and cytosine (GC) molecule. The excitation energies and oscillator strengths for the GC WC pair are given in Table 4, and a visualization of the charge density difference for selected excitations is given in Figure 5. Overall, CAM-B3LYP, LC-BLYP$_{α=0.0,β=1.0}$, and LC-BLYP$_{α=0.2,β=0.8}$ more accurately predict the excited state energies and oscillator strengths of this system compared to either M06-HF or B3LYP. In contrast to the excitations in the previous $π$-stacked complexes, we find *extreme* long-range charge-transfer excitations (i.e., CT$_{G→C}$ and Cn$_N$→$π$*) with very low $Λ$ values. The CT$_{G→C}$ transition has a $Λ$ value of 0.07, indicating a very low degree of orbital overlap and significant long-range charge transfer character. Surprisingly, CAM-B3LYP describes this transition extremely poorly and even predicts it as the *lowest-energy* excitation. In contrast, the other range-separated functionals with 100% asymptotic HF exchange predict the CT$_{G→C}$ transition quite accurately with errors of 0.15 and 0.07 eV for LC-BLYP$_{α=0.0,β=1.0}$ and LC-BLYP$_{α=0.2,β=0.8}$, respectively (in contrast to the very large 0.57 eV error for CAM-B3LYP). Since CAM-B3LYP is constructed with only 65% asymptotic HF exchange, the underestimation of the CT$_{G→C}$ excitation by CAM-B3LYP gives further evidence that a full 100% asymptotic HF exchange is essential for accurately describing these long-range excitations.

The Cn$_N$→π* transition is also a long-range charge-transfer excitation and has a Λ value of 0.13. Among all three of the range-separated functionals, the standard LC-BLYP$_{\alpha=0.0,\beta=1.0}$ has the highest error, and CAM-B3LYP closely reproduces the EOM-CCSD(T) benchmark excitation. The inclusion of some short-range HF exchange in LC-BLYP$_{\alpha=0.2,\beta=0.8}$ significantly improves upon the performance of the standard LC-BLYP functional, and brings the excitation energy much closer to the EOM-CCSD(T) benchmark. Both the Cn$_O$ → π* and Gn$_O$ → π* excitations have Λ values of 0.28 and 0.069, respectively, that are even lower than the CT$_{G \to C}$ transition. The errors from all of the functionals for the Cn$_O$ → π* and Gn$_O$ → π* excitations follows the same trend as the Cn$_N$ → π* transition discussed previously, giving further weight to the necessity of including short-range HF exchange. The Gπ → R + π* Rydberg excitation also has a high degree of charge transfer, with a Λ value of 0.21. This extended transition is only described accurately by range-separated functionals with some short-range HF exchange. Specifically, errors of ~0.06 eV are attained from both LC-BLYP$_{\alpha=0.2,\beta=0.8}$ and CAM-B3LYP, while a larger error of 0.11 eV is obtained from the LC-BLYP$_{\alpha=0.0,\beta=1.0}$ functional. The global hybrid functionals, in contrast, have large unsystematic errors for the Rydberg transition, with M06-HF severely overestimating the excitation by 0.68 eV, and B3LYP considerably underestimating the transition by 0.58 eV.

The ordering of the excited state energies relative to the EOM-CCSD(T) benchmarks is relatively good with the exception of a few outliers for the range-separated methods. The global hybrid methods, on the other hand, have several deviations in the energy ordering. For all of the methods except B3LYP, the 3(ππ*) transition is higher in energy than 4(ππ*), while the opposite is true in the EOM-CCSD(T) benchmark data. The LC-BLYP$_{\alpha=0.0,\beta=1.0}$ functional incorrectly places the 3(nπ*) transition lower in energy than the 5(ππ*) transition, an error that does not occur in the functionals that include some portion of short range exchange. In general, both of the global hybrids demonstrate extremely poor performance, with B3LYP significantly underestimating all excitations (especially the $CT_{G \to C}$ and $Cn_N \to \pi^*$ excitations) and M06-HF overestimating them. Overall, the LC-BLYP$_{\alpha=0.2, \beta=0.8}$ functional gives superior predictions for the GC WC pair, nearly halving the errors of both the standard LC-BLYP$_{\alpha=0.0, \beta=1.0}$ and CAM-B3LYP range-separated functionals.

**Conclusion**

In conclusion, we have examined an extensive and diverse set of electronic excitations in several DNA nucleobase monomers (adenine, cytosine, guanine, and thymine), stacked pair geometries (adenine-thymine and guanine-cytosine), and a canonical Watson-Crick base pair (guanine-cytosine). This diverse set of excitations comprises a total of 50 different transitions that include several $n \to \pi$ and $\pi \to \pi^*$ valence excitations, long-range charge-transfer excitations, and extended Rydberg transitions. Most importantly, the recent availability of high-level EOM-CCSD(T) benchmark calculations on these systems allows us to perform a stringent test of both short-range exchange and non-empirically tuned long-range exchange contributions for accurately predicting each of these various excitations.

From our results, we find that all the various range-separated functionals give an improved description of excited state properties, with an MAE of 0.11 – 0.16 eV, compared to the global hybrid functionals, which have a much larger MAE of 0.39 – 0.43 eV. In the case of the nucleobase monomers, both CAM-B3LYP and LC-BLYP$_{\alpha=0.2, \beta=0.8}$ perform slightly better than the standard LC-BLYP$_{\alpha=0.0, \beta=1.0}$ functional (both of the LC-BLYP methods contain a non-empirically tuned contribution of long-range exchange). We attribute this improved accuracy to the importance of short-range exchange (contained in both CAM-B3LYP and LC-BLYP$_{α=0.2,β=0.8}$) for reducing the self-interaction errors that are still present in these localized valence excitations. For both of the AT and GC stacked pair geometries, LC-BLYP$_{α=0.2,β=0.8}$ performs slightly better than both CAM-B3LYP and the standard LC-BLYP$_{α=0.0,β=1.0}$ functional. Finally, the GC Watson-Crick pair poses a serious challenge for TD-DFT methods, where we find both localized valence excitations as well as *extreme* long-range charge-transfer excitations with very small orbital overlaps. Among all the tested methods, the LC-BLYP$_{α=0.2,β=0.8}$ functional gives superior predictions (MAE = 0.08 eV and RMS Error = 0.10 eV), with errors that are considerably less than (*almost half*) the standard LC-BLYP$_{α=0.0,β=1.0}$ and CAM-B3LYP range-separated functionals. Based on these extensive results, we strongly recommend the use of the non-empirically tuned LC-BLYP$_{α=0.2,β=0.8}$ functional for simultaneously predicting all the diverse transition energies and properties in these various nucleobase complexes. In closing, the current study emphasizes the importance of *both* short-range exchange and a non-empirically tuned contribution of long-range exchange for predicting the diverse excitations in these challenging nucleobase systems. The improved accuracy and efficiency of these functionals allows further studies on large extended structures of nucleobases, where both accuracy and computational efficiency are critical for probing the electron dynamics and charge transport in these large systems.

**Acknowledgment**

We acknowledge the National Science Foundation for the use of supercomputing resources through the Extreme Science and Engineering Discovery Environment (XSEDE), Project no. TG-DMR140054.

**Supporting Information**

Reference Cartesian coordinates and electron density difference maps for all 50 excited states of the nucleobase structures. This information is available free of charge via the Internet at http://pubs.acs.org/.

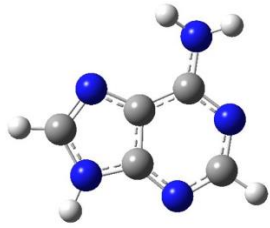 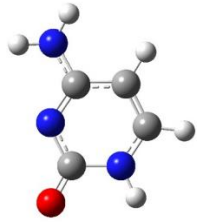 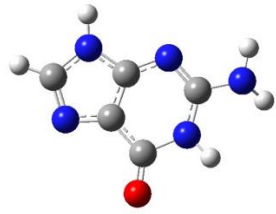 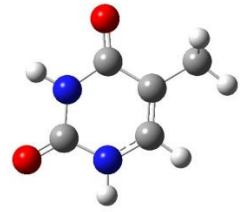

Adenine　　　　　　Cytosine　　　　　　Guanine　　　　　　Thymine

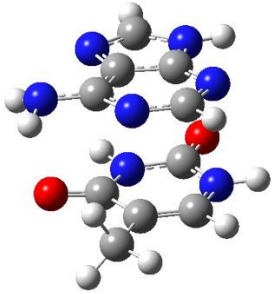 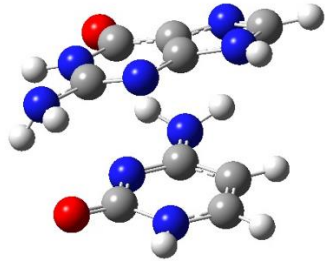 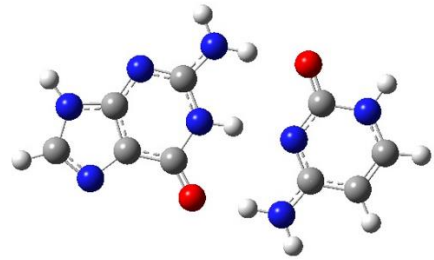

Adenine-Thymine Stack　　　Guanine-Cytosine Stack　　　Guanine-Cytosine Watson-Crick Pair

**Figure 1.** Molecular structures of the nucleobase monomers, stacked pairs, and Watson-Crick base pair.

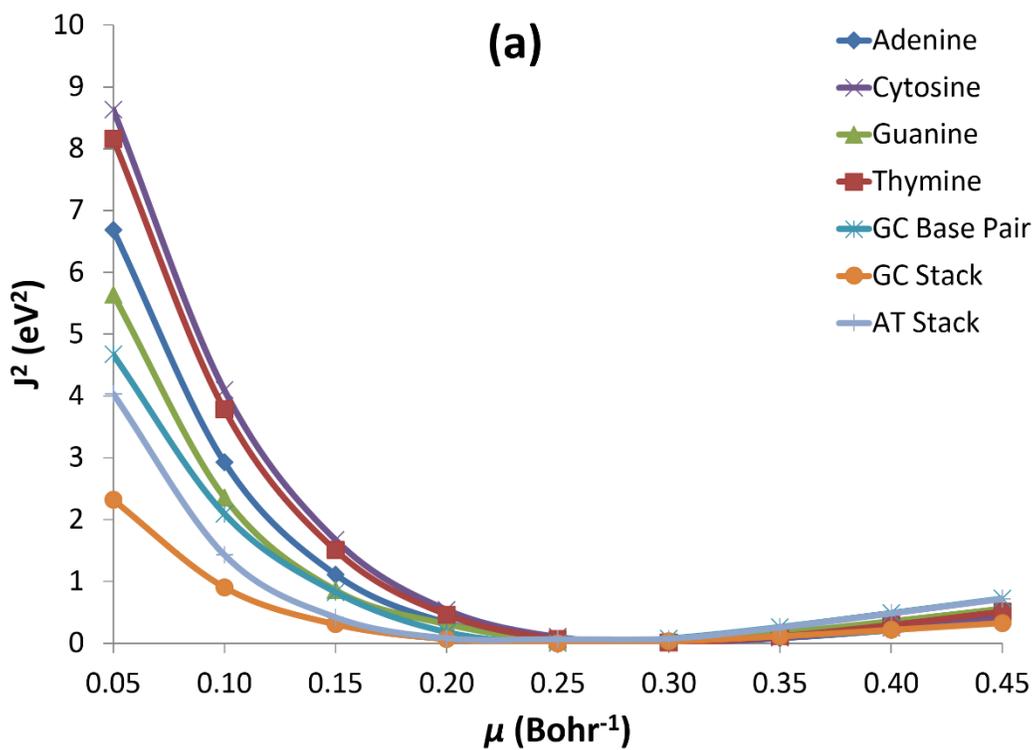

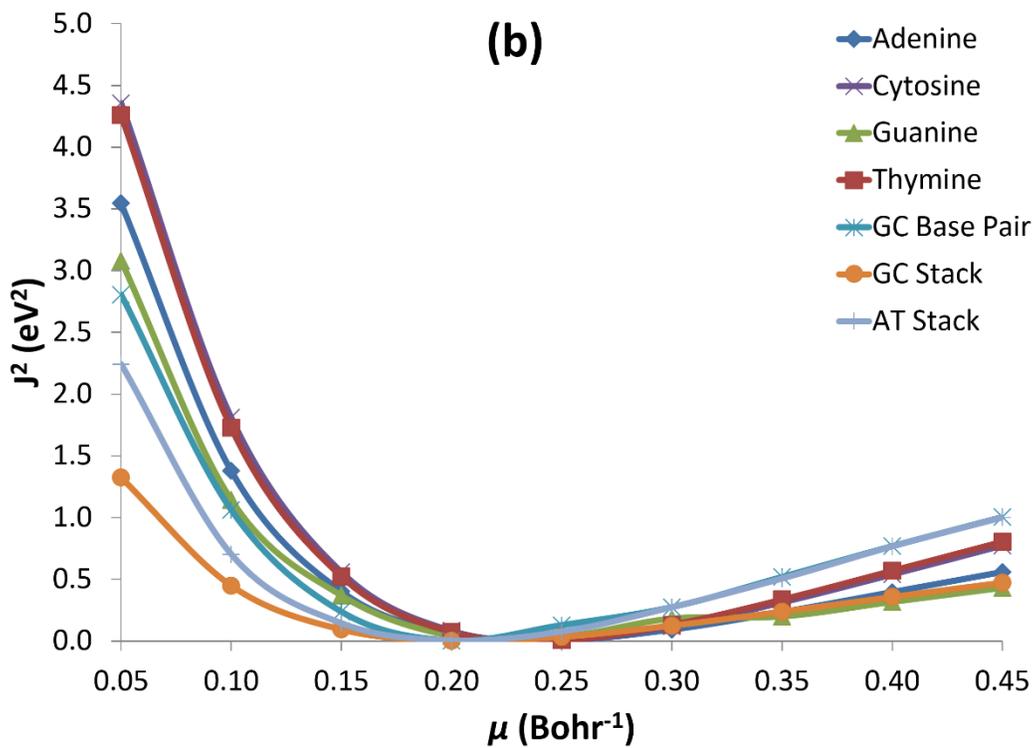

**Figure 2**: Plots of $J^2$ (eq. 3) as a function of $\mu$ for the various DNA bases and pairs using the TZVP basis with the (a) LC-BLYP$_{\alpha=0.0,\beta=1.0}$ and (b) LC-BLYP$_{\alpha=0.2,\beta=0.8}$ functionals.

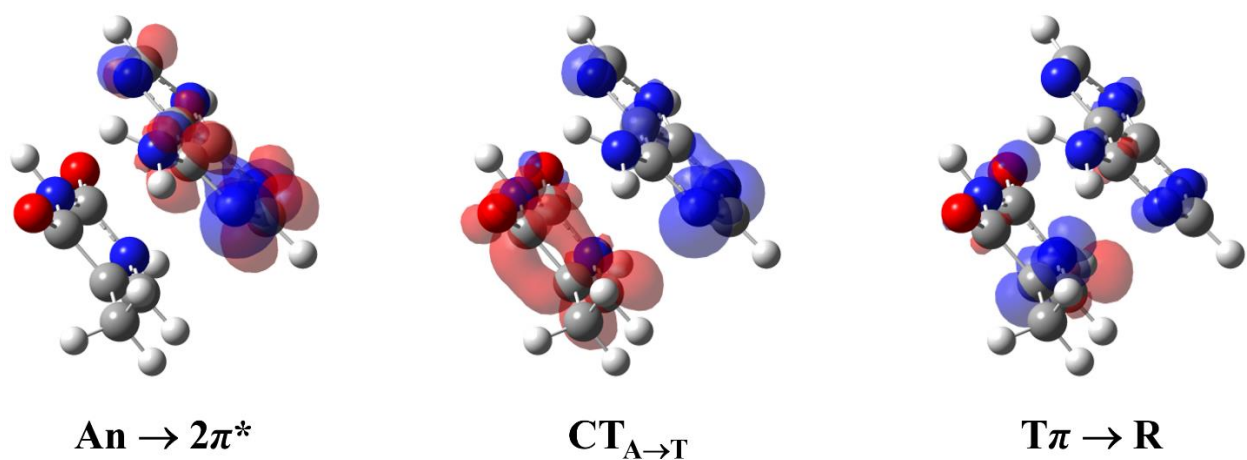

**Figure 3**. Electron density difference maps for (a) the An → $2\pi^*$ valence excitation, (b) the $CT_{A \to T}$ charge-transfer excitation and (c) the T$\pi$ → R Rydberg excitation in the adenine-thymine stacked pair. Red regions denote an accumulation of density upon electronic excitation, and blue regions represent a depletion of density upon excitation.

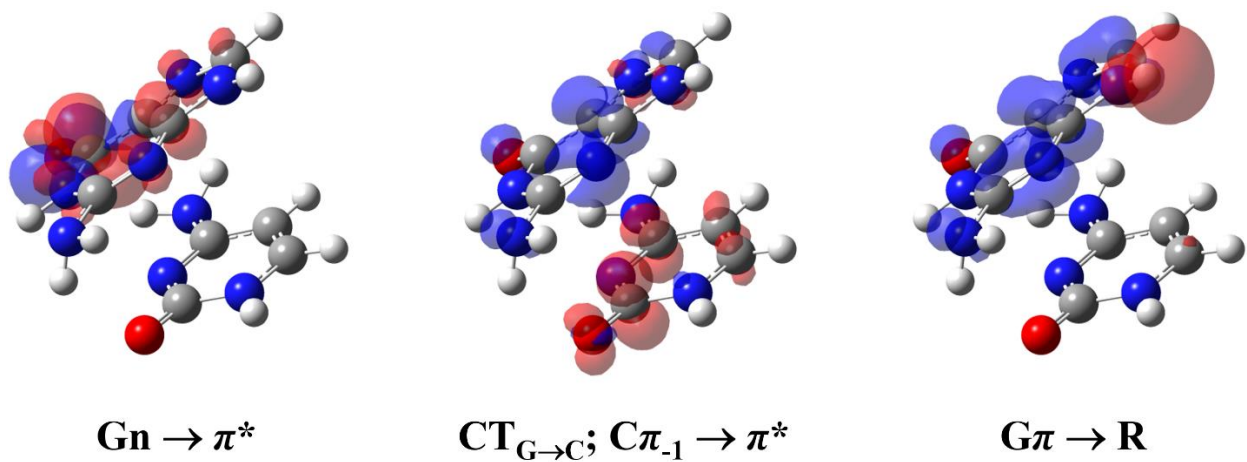

**Figure 4.** Electron density difference maps for (a) the Gn → π* valence excitation, (b) the $CT_{G \to C}$; $C\pi_{-1} \to \pi^*$ charge-transfer excitation, and (c) the Gπ → R Rydberg excitation in the guanine-cytosine stacked pair. Red regions denote an accumulation of density upon electronic excitation, and blue regions represent a depletion of density upon excitation.

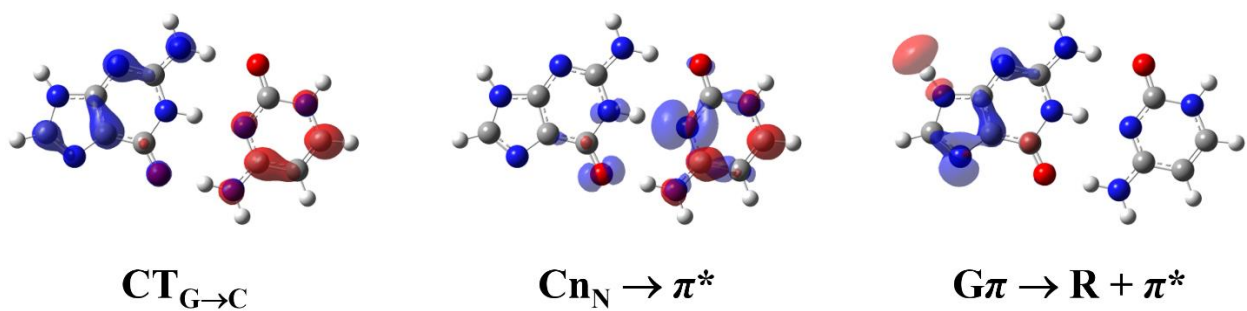

**Figure 5**. Electron density difference maps for (a) the CT$_{G \to C}$ charge transfer excitation, (b) the Cn$_N$ → $\pi^*$ valence excitation, and (c) the G$\pi$ → R + $\pi^*$ Rydberg excitation in the guanine-cytosine Watson-Crick base pair. Red regions denote an accumulation of density upon electronic excitation, and blue regions represent a depletion of density upon excitation.

**Table 1**: LC-BLYP/ TZVP optimal $\mu$ values for the DNA bases and pairs

|  | Optimal $\mu$ (Bohr$^{-1}$) | |
| --- | --- | --- |
|  | LC-BLYP$_{\alpha=0.0,\beta=1.0}$ | LC-BLYP$_{\alpha=0.2,\beta=0.8}$ |
| Adenine (A) | 0.288 | 0.236 |
| Cytosine (C) | 0.296 | 0.238 |
| Guanine (G) | 0.273 | 0.225 |
| Thymine (T) | 0.289 | 0.236 |
| GC Pair | 0.253 | 0.209 |
| GC Stack | 0.251 | 0.207 |
| AT Stack | 0.247 | 0.206 |

**Table 2**: Mean absolute errors, max errors, and root mean square errors in comparison to EOM-CCSD(T) benchmarks for the DNA monomers, stacked dimers, and WC pair.

|  | MAE | | | | |
|---|---|---|---|---|---|
|  | M06-HF | B3LYP | CAM-B3LYP | LC-BLYP$_{\alpha=0.0,\beta=1.0}$ | LC-BLYP$_{\alpha=0.2,\beta=0.8}$ |
| Adenine | 0.265 | 0.321 | 0.091 | 0.168 | 0.087 |
| Cytosine | 0.561 | 0.360 | 0.120 | 0.082 | 0.163 |
| Guanine | 0.395 | 0.383 | 0.053 | 0.091 | 0.042 |
| Thymine | 0.282 | 0.460 | 0.083 | 0.162 | 0.097 |
| AT Stack | 0.314 | 0.376 | 0.163 | 0.246 | 0.164 |
| GC Stack | 0.416 | 0.340 | 0.192 | 0.178 | 0.162 |
| GC Pair | 0.524 | 0.727 | 0.149 | 0.171 | 0.079 |
| Average MAE | 0.394 | 0.424 | 0.122 | 0.157 | 0.114 |
|  | Max Error | | | | |
|  | M06-HF | B3LYP | CAM-B3LYP | LC-BLYP$_{\alpha=0.0,\beta=1.0}$ | LC-BLYP$_{\alpha=0.2,\beta=0.8}$ |
| Adenine | 0.588 | 0.503 | 0.198 | 0.346 | 0.213 |
| Cytosine | 1.422 | 0.619 | 0.215 | 0.150 | 0.238 |
| Guanine | 0.609 | 0.638 | 0.085 | 0.168 | 0.073 |
| Thymine | 0.620 | 0.665 | 0.166 | 0.277 | 0.166 |
| AT Stack | 0.702 | 0.582 | 0.709 | 0.414 | 0.474 |
| GC Stack | 0.710 | 0.738 | 0.597 | 0.413 | 0.559 |
| GC Pair | 1.315 | 2.122 | 0.572 | 0.335 | 0.226 |
| Average Max Error | 0.852 | 0.838 | 0.363 | 0.300 | 0.279 |
|  | RMS Error | | | | |
|  | M06-HF | B3LYP | CAM-B3LYP | LC-BLYP$_{\alpha=0.0,\beta=1.0}$ | LC-BLYP$_{\alpha=0.2,\beta=0.8}$ |
| Adenine | 0.327 | 0.357 | 0.113 | 0.189 | 0.114 |
| Cytosine | 0.721 | 0.396 | 0.134 | 0.095 | 0.174 |
| Guanine | 0.414 | 0.428 | 0.060 | 0.111 | 0.050 |
| Thymine | 0.364 | 0.480 | 0.100 | 0.187 | 0.106 |
| AT Stack | 0.379 | 0.398 | 0.257 | 0.279 | 0.200 |
| GC Stack | 0.448 | 0.416 | 0.268 | 0.216 | 0.233 |
| GC Pair | 0.597 | 0.926 | 0.221 | 0.204 | 0.101 |
| Average RMS Error | 0.464 | 0.486 | 0.165 | 0.183 | 0.140 |

**Table 3**: Excitation energies (ΔE in eV), oscillator strengths (*f* in au), and the lambda diagnostic (Λ) of the DNA nucleobase monomers calculated by various methods.

| | | \multicolumn{13}{c|}{Adenine} | |
|---|---|---|---|---|---|---|---|---|---|---|---|---|---|---|
| | | EOM-CCSD[a] | | EOM-CCSD(T)[a] | M06-HF | | B3LYP | | CAM-B3LYP | | LC-BLYP$_{\alpha=0.0,\beta=1.0}$ | | LC-BLYP$_{\alpha=0.2,\beta=0.8}$ | | |
| type | transition | ΔE | *f* | ΔE | ΔE | *f* | ΔE | *f* | ΔE | *f* | ΔE | *f* | ΔE | *f* | Λ |
| 1(ππ*) | Aπ → π* | 5.39 | 0.000 | 5.17 | 5.59 | 0.005 | 4.91 | 0.002 | 5.31 | 0.001 | 5.07 | 0.001 | 5.31 | 0.001 | 0.435 |
| 2(ππ*) | Aπ → 2π* | 5.66 | 0.297 | 5.48 | 5.68 | 0.306 | 5.05 | 0.201 | 5.38 | 0.280 | 5.34 | 0.269 | 5.41 | 0.283 | 0.796 |
| 2(nπ*) | An → 2π* | 5.52 | 0.006 | 5.30 | 5.89 | 0.049 | 5.29 | 0.036 | 5.50 | 0.009 | 5.40 | 0.018 | 5.51 | 0.012 | 0.711 |
| 3(nπ*) | An → π* | 6.13 | 0.005 | 5.93 | 5.97 | 0.002 | 5.57 | 0.003 | 5.92 | 0.003 | 5.73 | 0.004 | 5.93 | 0.003 | 0.487 |
| 4(nπ*) | An$_{-1}$ → 2π* | 6.56 | 0.007 | 6.35 | 6.42 | 0.007 | 5.85 | 0.002 | 6.27 | 0.004 | 6.00 | 0.004 | 6.26 | 0.004 | 0.444 |
| 5(ππ*) | Aπ → π*[b] | 6.82 | 0.223 | 6.68 | 6.95 | 0.454 | 6.32 | 0.209 | 6.66 | 0.428 | 6.56 | 0.408 | 6.67 | 0.390 | 0.667 |

| | | \multicolumn{13}{c|}{Cytosine} | |
|---|---|---|---|---|---|---|---|---|---|---|---|---|---|---|
| | | EOM-CCSD[a] | | EOM-CCSD(T)[a] | M06-HF | | B3LYP | | CAM-B3LYP | | LC-BLYP$_{\alpha=0.0,\beta=1.0}$ | | LC-BLYP$_{\alpha=0.2,\beta=0.8}$ | | |
| type | transition | ΔE | *f* | ΔE | ΔE | *f* | ΔE | *f* | ΔE | *f* | ΔE | *f* | ΔE | *f* | Λ |
| 1(ππ*) | Cπ → π* | 4.98 | 0.070 | 4.76 | 5.21 | 0.088 | 4.65 | 0.036 | 4.97 | 0.068 | 4.91 | 0.068 | 5.00 | 0.072 | 0.609 |
| 1(nπ*) | Cn$_N$ → π* | 5.39 | 0.004 | 5.18 | 5.30 | 0.014 | 4.77 | 0.008 | 5.22 | 0.003 | 5.07 | 0.003 | 5.23 | 0.003 | 0.347 |
| 3(ππ*) | Cπ$_{-1}$ → π* | 6.09 | 0.160 | 5.84 | 6.37 | 0.252 | 5.51 | 0.083 | 6.00 | 0.117 | 5.93 | 0.122 | 6.04 | 0.122 | 0.636 |
| 3(nπ*) | Cn$_o$ → 2π* | 5.97 | 0.008 | 5.73 | 5.45 | 0.009 | 5.11 | 0.003 | 5.83 | 0.003 | 5.71 | 0.001 | 5.89 | 0.003 | 0.452 |
| 4(nπ*) | Cn$_o$ → π* | 6.42 | 0.003 | 6.02 | 7.44 | 0.001 | 5.69 | 0.001 | 6.10 | 0.005 | 5.99 | 0.007 | 6.19 | 0.004 | 0.394 |

| | | \multicolumn{13}{c|}{Guanine} | |
|---|---|---|---|---|---|---|---|---|---|---|---|---|---|---|
| | | EOM-CCSD[a] | | EOM-CCSD(T)[a] | M06-HF | | B3LYP | | CAM-B3LYP | | LC-BLYP$_{\alpha=0.0,\beta=1.0}$ | | LC-BLYP$_{\alpha=0.2,\beta=0.8}$ | | |
| type | transition | ΔE | *f* | ΔE | ΔE | *f* | ΔE | *f* | ΔE | *f* | ΔE | *f* | ΔE | *F* | Λ |
| 2(ππ*) | Gπ → π* | 5.32 | 0.184 | 5.12 | 5.45 | 0.211 | 4.92 | 0.161 | 5.18 | 0.181 | 5.09 | 0.167 | 5.18 | 0.178 | 0.722 |
| 2(nπ*) | Gn → π* | 5.67 | 0.002 | 5.53 | 4.92 | 0.006 | 5.34 | 0.069 | 5.54 | 0.009 | 5.39 | 0.005 | 5.54 | 0.008 | 0.531 |
| 4(ππ*) | Gπ → 2π* | 5.99 | 0.318 | 5.77 | 6.08 | 0.421 | 5.27 | 0.140 | 5.69 | 0.288 | 5.60 | 0.287 | 5.70 | 0.296 | 0.647 |
| 1(πR) | Gπ → R | 6.27 | 0.004 | 6.11 | 6.44 | 0.005 | 5.47 | 0.007 | 6.17 | 0.003 | 6.09 | 0.004 | 6.13 | 0.003 | 0.326 |

| | | \multicolumn{13}{c|}{Thymine} | |
|---|---|---|---|---|---|---|---|---|---|---|---|---|---|---|
| | | EOM-CCSD[a] | | EOM-CCSD(T)[a] | M06-HF | | B3LYP | | CAM-B3LYP | | LC-BLYP$_{\alpha=0.0,\beta=1.0}$ | | LC-BLYP$_{\alpha=0.2,\beta=0.8}$ | | |
| type | transition | ΔE | *f* | ΔE | ΔE | *f* | ΔE | *f* | ΔE | *f* | ΔE | *f* | ΔE | *F* | Λ |
| 1(nπ*) | Tn → π* | 5.20 | 0.000 | 5.03 | 4.66 | 0.000 | 4.74 | 0.000 | 5.09 | 0.000 | 4.99 | 0.000 | 5.12 | 0.000 | 0.406 |

| | | | | | | | | | | | | | | |
|---|---|---|---|---|---|---|---|---|---|---|---|---|---|---|
| 2(ππ*) | Tπ → π* | 5.65 | 0.235 | 5.47 | 5.54 | 0.267 | 5.04 | 0.136 | 5.30 | 0.192 | 5.19 | 0.177 | 5.30 | 0.195 | 0.728 |
| 5(nπ*) | $T_{n-1}$ → 2π* | 6.65 | 0.000 | 6.53 | 5.91 | 0.000 | 5.87 | 0.000 | 6.44 | 0.000 | 6.31 | 0.000 | 6.47 | 0.000 | 0.358 |
| 1(πR) | Tπ → R | 6.78 | 0.000 | 6.67 | 6.74 | 0.000 | 6.21 | 0.000 | 6.66 | 0.000 | 6.56 | 0.000 | 6.60 | 0.000 | 0.311 |

[a]From Szalay et al.[17]

**Table 4.** Excitation energies (ΔE in eV), oscillator strengths (*f* in au), and the lambda diagnostic (Λ) of the adenine-thymine and guanine-cytosine stacked dimers and the guanine-cytosine Watson-Crick base pair calculated by various methods.

| | | AT Stack | | | | | | | | | | | | | |
|---|---|---|---|---|---|---|---|---|---|---|---|---|---|---|---|
| | | EOM-CCSD[a] | | EOM-CCSD(T)[a] | M06-HF | | B3LYP | | CAM-B3LYP | | LC-BLYP$_{\alpha=0.0,\beta=1.0}$ | | LC-BLYP$_{\alpha=0.2,\beta=0.8}$ | | |
| type | transition | ΔE | *f* | ΔE | ΔE | *f* | ΔE | *f* | ΔE | *f* | ΔE | *f* | ΔE | *f* | Λ |
| 1(nπ*) | Tn → π* | 5.22 | 0.000 | 5.04 | 4.72 | 0.000 | 4.79 | 0.000 | 5.12 | 0.009 | 4.88 | 0.002 | 5.08 | 0.008 | 0.408 |
| 1(ππ*) | Aπ → π* | 5.36 | 0.000 | 5.13 | 5.81 | 0.021 | 5.24 | 0.047 | 5.46 | 0.061 | 5.23 | 0.031 | 5.38 | 0.048 | 0.668 |
| 2(ππ*) | Tπ → π*; Aπ → 2π* | 5.42 | 0.024 | 5.23 | 5.39 | 0.023 | 4.88 | 0.021 | 5.17 | 0.022 | 4.98 | 0.023 | 5.12 | 0.020 | 0.617 |
| 2(nπ*) | An → 2π* | 5.51 | 0.004 | 5.31 | 5.59 | 0.031 | 4.89 | 0.004 | 5.30 | 0.047 | 4.90 | 0.001 | 5.22 | 0.011 | 0.436 |
| 3(ππ*) | Tπ → π*; Aπ → 2π* | 5.64 | 0.339 | 5.46 | 5.64 | 0.360 | 5.04 | 0.171 | 5.32 | 0.127 | 5.15 | 0.179 | 5.28 | 0.197 | 0.720 |
| 3(nπ*) | An → π* | 6.14 | 0.002 | 5.94 | 5.99 | 0.002 | 5.58 | 0.002 | 5.94 | 0.004 | 5.58 | 0.003 | 5.86 | 0.003 | 0.454 |
| 4(ππ*) | CT$_{A→T}$ | 6.17 | 0.052 | 5.88 | 6.58 | 0.074 | 5.52 | 0.011 | 6.59 | 0.048 | 6.00 | 0.001 | 6.35 | 0.012 | 0.371 |
| 4(nπ*) | An$_{-1}$ → 2π* | 6.49 | 0.005 | 6.27 | 6.41 | 0.005 | 5.81 | 0.001 | 6.25 | 0.005 | 5.87 | 0.009 | 6.15 | 0.006 | 0.307 |
| 5(ππ*) | Aπ → π* | 6.60 | 0.090 | 6.37 | 6.65 | 0.062 | 6.10 | 0.017 | 6.63 | 0.122 | 6.36 | 0.188 | 6.60 | 0.111 | 0.561 |
| 5(nπ*) | T$_{n-1}$ → 2π* | 6.67 | 0.008 | 6.54 | 6.03 | 0.000 | 5.96 | 0.001 | 6.45 | 0.004 | 6.16 | 0.006 | 6.43 | 0.033 | 0.372 |
| 1(πR) | Tπ → R | 6.77 | 0.033 | 6.63 | 6.78 | 0.030 | 6.07 | 0.008 | 6.72 | 0.152 | 6.43 | 0.038 | 6.50 | 0.089 | 0.306 |
| | | GC Stack | | | | | | | | | | | | | |
| | | EOM-CCSD[a] | | EOM-CCSD (T)[a] | M06-HF | | B3LYP | | CAM-B3LYP | | LC-BLYP$_{\alpha=0.0,\beta=1.0}$ | | LC-BLYP$_{\alpha=0.2,\beta=0.8}$ | | |
| type | transition | ΔE | *f* | ΔE | ΔE | *f* | ΔE | *f* | ΔE | *f* | ΔE | *f* | ΔE | *f* | Λ |
| 1(ππ*) | Cπ → π* | 5.07 | 0.045 | 4.83 | 5.29 | 0.049 | 4.73 | 0.018 | 5.08 | 0.038 | 4.84 | 0.023 | 4.97 | 0.016 | 0.620 |
| 2(ππ*) | Gπ → π* | 5.25 | 0.044 | 5.04 | 5.39 | 0.045 | 4.95 | 0.072 | 5.24 | 0.098 | 4.97 | 0.038 | 5.19 | 0.039 | 0.696 |
| 1(nπ*) | Cn$_N$ → π* | 5.41 | 0.035 | 5.20 | 5.45 | 0.066 | 4.87 | 0.009 | 5.26 | 0.010 | 4.88 | 0.006 | 5.23 | 0.066 | 0.453 |
| 2(nπ*) | Gn → π* | 5.66 | 0.004 | 5.51 | 5.02 | 0.001 | 5.24 | 0.003 | 5.52 | 0.001 | 5.29 | 0.002 | 5.48 | 0.002 | 0.397 |
| 3(ππ*) | CT$_{G→C}$; Cπ$_{-1}$ → π* | 5.69 | 0.150 | 5.42 | 6.07 | 0.177 | 5.38 | 0.023 | 5.83 | 0.098 | 5.57 | 0.078 | 5.76 | 0.088 | 0.323 |
| 4(ππ*) | Gπ → 2π* | 5.80 | 0.208 | 5.57 | 5.88 | 0.300 | 5.06 | 0.151 | 5.50 | 0.223 | 5.34 | 0.199 | 5.46 | 0.217 | 0.727 |
| 5(ππ*) | CT$_{G→C}$; Cπ$_{-1}$ → π* | 5.99 | 0.068 | 5.71 | 6.42 | 0.043 | 5.40 | 0.040 | 6.31 | 0.029 | 6.12 | 0.046 | 6.27 | 0.037 | 0.358 |
| 3(nπ*) | Cn$_o$ → 2π* | 6.15 | 0.003 | 5.94 | 5.73 | 0.002 | 5.20 | 0.024 | 6.03 | 0.003 | 5.84 | 0.006 | 5.98 | 0.002 | 0.439 |
| 1(πR) | Gπ → R | 6.49 | 0.002 | 6.32 | 6.65 | 0.012 | 5.64 | 0.001 | 6.36 | 0.000 | 6.23 | 0.009 | 6.25 | 0.001 | 0.245 |
| | | GC Watson-Crick Pair | | | | | | | | | | | | | |
| | | EOM-CCSD[a] | | EOM-CCSD (T)[a] | M06-HF | | B3LYP | | CAM-B3LYP | | LC-BLYP$_{\alpha=0.0,\beta=1.0}$ | | LC-BLYP$_{\alpha=0.2,\beta=0.8}$ | | |

| type | transition | ΔE | f | ΔE | ΔE | f | ΔE | f | ΔE | f | ΔE | f | ΔE | f | Λ |
|---|---|---|---|---|---|---|---|---|---|---|---|---|---|---|---|
| 1(ππ*) | Gπ → π* | 5.07 | 0.065 | 4.85 | 5.21 | 0.097 | 4.76 | 0.101 | 5.10 | 0.070 | 4.81 | 0.071 | 4.93 | 0.076 | 0.694 |
| 2(ππ*) | Cπ → π* | 5.17 | 0.100 | 4.92 | 5.39 | 0.199 | 4.77 | 0.028 | 5.14 | 0.102 | 4.94 | 0.082 | 5.10 | 0.097 | 0.619 |
| 3(ππ*) | Cπ₋₁ → π* | 5.64 | 0.303 | 5.37 | 6.03 | 0.191 | 5.00 | 0.037 | 5.65 | 0.120 | 5.42 | 0.102 | 5.60 | 0.112 | 0.533 |
| 4(ππ*) | Gπ → 2π* | 5.71 | 0.411 | 5.48 | 5.81 | 0.579 | 5.14 | 0.278 | 5.47 | 0.431 | 5.34 | 0.414 | 5.45 | 0.428 | 0.690 |
| 5(ππ*) | CT$_{G→C}$ | 5.81 | 0.013 | 5.36 | 6.68 | 0.001 | 3.24 | 0.002 | 4.79 | 0.024 | 5.21 | 0.016 | 5.29 | 0.022 | 0.069 |
| 1(nπ*) | Cn$_N$ → π* | 5.87 | 0.001 | 5.65 | 6.10 | 0.001 | 4.69 | 0.000 | 5.62 | 0.001 | 5.32 | 0.001 | 5.58 | 0.001 | 0.134 |
| 2(nπ*) | Gn$_O$ → π* | 5.95 | 0.000 | 5.76 | 5.48 | 0.000 | 5.50 | 0.000 | 5.79 | 0.000 | 5.56 | 0.000 | 5.77 | 0.000 | 0.480 |
| 5(ππ*) | Gπ → R + π* | 6.36 | 0.000 | 6.20 | 6.88 | 0.004 | 5.62 | 0.000 | 6.27 | 0.001 | 6.09 | 0.000 | 6.14 | 0.001 | 0.207 |
| 3(nπ*) | Cn$_O$ → 2π* | 6.43 | 0.000 | 6.27 | 5.97 | 0.000 | 5.27 | 0.000 | 6.31 | 0.000 | 5.94 | 0.000 | 6.29 | 0.000 | 0.239 |
| 4(nπ*) | Cn$_O$ → π* | 6.60 | 0.003 | 6.42 | 6.72 | 0.289 | 5.57 | 0.001 | 6.40 | 0.000 | 6.09 | 0.000 | 6.34 | 0.000 | 0.280 |
| 6(ππ*) | Cπ → 2π* | 6.71 | 0.226 | 6.50 | 7.11 | 0.172 | 5.23 | 0.087 | 6.61 | 0.158 | 6.33 | 0.101 | 6.56 | 0.128 | 0.432 |

[a]From Szalay et al.[17]

TOC Graphic:

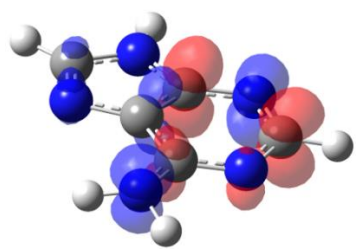 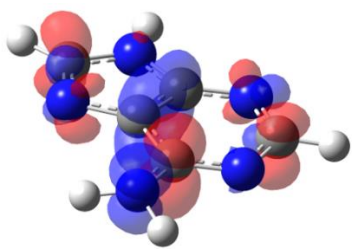 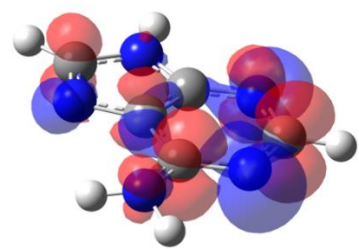

**An → 2π\***          **Aπ → 2π\***         **Aπ → π\***

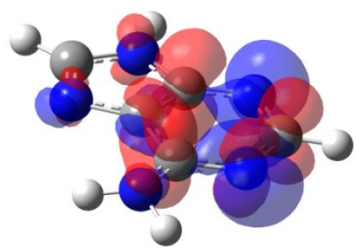 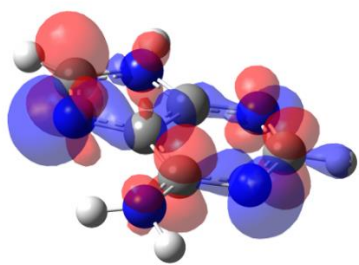 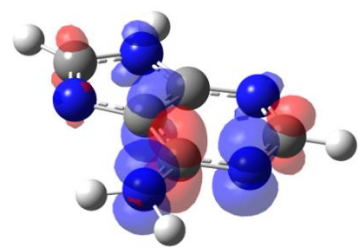

**An → π\***         **An$_{-1}$ → 2π\***         **Aπ → π\*b**